# Optospintronics in graphene via proximity coupling

Ahmet Avsar[1*], Dmitrii Unuchek[1], Jiawei Liu[2], Oriol Lopez Sanchez[1], Kenji Watanabe[3], Takashi Taniguchi[3], Barbaros Özyilmaz[2], Andras Kis[1*]

[1] Electrical Engineering Institute and Institute of Materials Science and Engineering, École Polytechnique Fédérale de Lausanne (EPFL), Lausanne CH 1015, Switzerland

[2] Centre for Advanced 2D Materials, National University of Singapore, Singapore 117542, Singapore

[3] National Institute for Materials Science, 1-1 Namiki, Tsukuba 305-0044, Japan

**The observation of micron size spin relaxation makes graphene a promising material for applications in spintronics requiring long-distance spin communication[1]. However, spin dependent scatterings at the contact/graphene interfaces affect the spin injection efficiencies and hence prevent the material from achieving its full potential. While this major issue could be eliminated by nondestructive direct optical spin injection schemes, graphene's intrinsically low spin-orbit coupling strength and optical absorption place an obstacle in their realization[2]. We overcome this challenge by creating sharp artificial interfaces between graphene and $WSe_2$ monolayers. Application of a circularly polarized light activates the spin-polarized charge carriers in the $WSe_2$ layer due to its spin-coupled valley-selective absorption[3]. These carriers diffuse into the superjacent graphene layer, transport over a 3.5 μm distance, and are finally detected electrically using h-BN/Co contacts in a non-local geometry. Polarization-dependent measurements confirm the spin origin of the non-local signal.**

Spintronics has been proposed for applications in logic devices as a complement or even an alternative to devices based on the charge degree of freedom[4,5]. Searching for the ideal material that can transport spin-dependent currents beyond micron size distances has been one of the main focuses of spintronics research[5]. In this respect, graphene is a unique material due to its low spin-orbit coupling[6], negligible hyperfine interaction[7], large Fermi velocity[1] and very high electronic mobility[8]. Indeed, it exhibits the longest spin relaxation length at room temperature measured by magnetoresistance electrical measurements using ferromagnetic electrodes[9]. However, the notorious conductivity mismatch issue in this measurement scheme[10] limits the effective spin injection severely affecting the coherent spin transport over long distances[1]. Despite the use of tunnel barriers such as thin oxide layers[11,12] and pin-hole free boron nitride (BN) crystals between graphene and ferromagnetic electrodes[13], the striking difference between theoretically predicted and experimentally observed spin relaxation lengths still persists[1] and is waiting to be addressed. Nondestructive optical spin injection schemes could be an appealing alternative. However, the absence of sufficient spin-orbit coupling and weak optical absorption of graphene poses challenges for their implementation[2].

In contrast, transition metal dichalcogenides (TMDCs) exhibit strong light absorption even in their monolayers[14,15]. Monolayer TMDCs also have unique spin-valley physics which has captured the interest of the solid state physics community[16,3,17]. Due to the broken spin degeneracy and the time-reversal symmetry, the spin and valley degrees of freedom are coupled in a such way that excitation by opposite handedness leads to preferential population of the K or K' valley with a defined spin orientation. Among TMDCs, tungsten diselenide ($WSe_2$) gets special attention due to the achievement of a valley polarization value close to unity and large external quantum efficiencies[18,19]. It also has the highest valance band splitting of ~ 450 meV[20].

As proposed by M. Gmitra & J. Fabian[2], the generation of spin polarized charge carriers by using its spin coupled valley selective absorption property[21] could allow inducing spin dependent currents in the superjacent graphene layer through a tunneling process, without the need for a ferromagnetic spin injector.

Towards this, we fabricate heterostructure devices consisting of monolayer $WSe_2$, graphene and BN (Figure 1a). Our fabrication process starts with the mechanical exfoliation of monolayer graphene on a conventional $SiO_2$ (~ 270 nm)/Si wafer. Next, graphene is partially covered with monolayer $WSe_2$ by utilizing the dry transfer method described in reference [8]. Finally, a three-layer BN crystal is similarly transferred by targeting the uncovered region of graphene. Optical images for one of our typical heterostructure devices at different fabrication steps are shown in Figure 1b. Before the metallization process, the heterostructure is annealed at 250 ºC under high vacuum conditions (~ 5 x $10^{-7}$ Torr) for 6 hours. This process results in a cleaner interface between two-dimensional (2D) materials by removing the transfer-related residues. Electrode masks are prepared by utilizing the standard electron beam lithography technique. Device fabrication is completed by forming the Co / Ti (35 nm / 10 nm) electrodes. Deposition rate for both materials is ~ 0.5 Å/sec and the Ti layer serves as a capping layer to prevent the oxidation of ferromagnetic Co electrodes. All-electrical spin transport measurements are performed with a standard AC lock-in technique at low frequencies and at fixed current bias in the four-terminal non-local spin valve geometry as a function of in-plane and out of plane magnetic field ($B_\parallel$ and $B_\perp$). For opto-electrical measurements, the monolayer $WSe_2$ flake is resonantly excited at 720 nm (1.72 eV, black arrow in Figure 3b) using a supercontinuum laser with a maximal incident light intensity of 190 μW at 4K. In order to detect the non-local voltage signal, the lock-in technique is combined with a photoelastic modulator (PEM), acting as a variable waveplate. The linear polarizer in front of PEM is used to control the angle between incident light polarization and the optical axis of the modulator. This way, in the case of quarter-wave modulation while using the fundamental frequency of PEM as the lock-in reference signal (50 kHz), the maximal amplitude of the signal should match the angle θ of + (-) 45º that corresponds to the right-to-left (RL) (left-to-right (LR)) modulation. During polarization dependent measurements, the ellipticity of the light is modified by changing the polarization angle θ. This decreases the signal amplitude, reaching the lock-in noise floor when incident light polarization is parallel to the PEM optical axis. For the experiment with linear light modulation, half-wave configuration of PEM was used with double frequency (100 kHz) as a reference signal. The details of the optospintronics measurement setup are shown in Supplementary Information. Here, we represent results obtained in two different devices, labelled as device A and device B. Unless otherwise stated, the results shown are from device A.

Prior to any optical measurements, we first characterize a graphene-based spin valve device with trilayer BN used as a tunnel barrier. Figure 2a shows the device resistance as a function of back-gate voltage ($V_{BG}$). Our device characteristic shows the typical ambipolar field effect behavior. The charge neutrality point is observed at negative $V_{BG}$ values which indicates the weak n-doped nature of graphene. We extract an electron mobility of ~ 5,500 $cm^2$/Vs at ~ 1 x $10^{12}$ $cm^{-2}$ carrier concentration. At low bias range, we observe a nearly linear I-V relation (Figure 2a-Inset). As the next step, we perform all-electrical spin injection, transport and detection measurement in a non-local geometry (Figure 2b-Inset). For this, we apply a fixed current of 5 µA between electrodes *1&2* and record the non-local voltage between electrodes *3&4* while sweeping the in-plane magnetic field $B_I$. This changes the relative polarization orientations of the injector (*2*) and detector (*3*) electrodes and induces a non-local spin signal of ~ 0.2 Ω. In order to determine the spin polarization ($P$) of electrodes (Co/3 layers of BN) which will be later employed for detecting the optically generated spin signal, we perform conventional Hanle precession measurements[22]. Here, the non-local signal is recorded while the out-of-plane magnetic field $B_\perp$ is swept in the range of ±150 mT. Since the spin-dependent current precesses along the field, the signal decreases (increases) for the parallel (antiparallel) configuration as the strength of the $B_\perp$ is increased. The resulting signal can be fitted with the solution of the Bloch equation[11],

$$R_{NL} \sim \int_0^\infty \frac{1}{\sqrt{4\pi D_s t}} exp\left(\frac{-L^2}{4D_s t}\right) exp\left(\frac{-t}{\tau_S}\right) \cos(w_L t) dt$$

where $L \approx 4$ μm is the center-to-center separation between the injector and detector electrodes and $w_L$ is the Larmor frequency. This gives a spin relaxation time of $\tau_S \approx 184 \pm 1$ ps, a spin diffusion constant of $D_S \approx 0.138$ m$^2$/s, and hence, a spin relaxation length of $\lambda_S \approx 5$ μm. Spin polarization value can be calculated by[23]

$$P = \left(\frac{2\Delta R w \sigma}{\lambda_S} e^{(L/\lambda_S)}\right)^{0.5}$$

where, $w$ and $\sigma$ are the width and conductivity of graphene, respectively. By inserting the $\lambda s$ extracted from spin precession measurements, we calculate $P$ to be ~ 0.6 %. The spin parameters extracted above strongly depend on the $V_{BG}$: the $P$ value is enhanced up to 2.85% at high $V_{BG}$ values (See Supplementary Information). We consistently observe a very similar response at all measured junctions within this device which indicates the large size uniformity of three-layer thick BN. Note that these spin polarization values are an order of magnitude smaller than for the best tunnel barriers ever created for graphene[24]. However, they are comparable to the values obtained using oxide based tunnel barriers[25] and reliable enough for detecting the optically injected spin currents.

Next, we characterize our WSe$_2$-graphene-BN heterostructure device. Figure 3a shows the V$_{BG}$ dependence of graphene resistance which is similarly measured by using trilayer BN/Co electrodes. Similarly to the data shown in Figure 2a, we observe ambipolar characteristic with a weak n-type doping. The corresponding I-V characteristic is also linear at low bias range. These results are consistent with the device performance shown in Figure 2a. This suggests that our electrode could serve as a spin detector. Next, we optically characterize the WSe$_2$ flake in our device. In addition to the flake optical contrast, we confirm the monolayer nature of WSe$_2$ by photoluminescence measurements using a 488 nm blue laser diode with low incident power of 40 µW. As shown in Figure 3b, we observe a strong peak X$^o$ near 1.66 eV corresponding to the excitonic resonance in a monolayer flake. We can also distinguish the lower energy X$^{'}$ peak associated with the trion as well as localized exciton emission[26] (See Supplementary Information).

Now we turn our attention to the optical spin injection aspect of our study. In order to assure the full out-of-plane direction magnetization of Co electrodes, we first apply B$_\perp$ = 2T and then set the field to B = 0[27]. Next, we focus the laser beam under quarter-wave modulation on the device and detect the generated non-local signal electrically in a non-local geometry. As shown in Figure 3c, we do not observe any significant non-local signal while the light spot is parked on graphene. In contrast, once the spot is placed on the WSe$_2$/graphene heterostructure, we observe a sudden increase in the non-local voltage reaching 1 µV, even though the laser beam is much further away from the detector electrodes compared to the initial case with the laser spot on graphene. The signal returns back to its initial value of ~ 0.1 µV when the spot is placed back on top of the graphene region. This measurement suggests that the measured signal is not due to spurious effects such as laser heating etc. The non-local origin of the signal is confirmed by the length-dependent measurement. As shown in Figure 3d, the magnitude of the signal decreases ~ 25% from the initial value if the electrode 2 is utilized as the detector which is ~ 3.7 μm far away from the graphene/WSe$_2$ interface. This is expected within the spin transport theory as the spin dephases more while it travels a longer distance and therefore the measured signal amplitude decreases[4]. It is also worth mentioning that the measured signal has a weak dependence on the location of the laser spot (See Supplementary Information). This could be related to the local interface homogeneity.

Figure 4a shows the spatial map of the non-local voltage signal measured in device B at $V_{BG} = 0$ V. Similar to the device A (Figure 3c), we detect the non-local voltage if only the laser spot is parked on the monolayer WSe$_2$. As we move the spot on bilayer WSe$_2$ region, we observe a sudden suppression of the non-local voltage and it is undetectable. This is indeed expected as the inversion symmetry is restored in bilayer devices and shows that the origin of the signal is the spin-coupled, valley-selective absorption.

In order to further confirm the spin origin of this signal, we measure the non-local signal under quarter-wave ($\lambda/4$) and half-wave ($\lambda/2$) modulations of incident light. Only the former modulation should result in the spin-dependent signal as the activation of a specific valley is only possible with the circularly polarized light, while the half-wave modulation does not meet this requirement. If the origin of the signal were not spin dependent, we would observe the same response under both modulations. Figure 4b shows the time dependence of the non-local voltage measured under $\lambda/4$, followed by $\lambda/2$ modulation. We observe a non-local signal of ~ 0.62 μV for the case of $\lambda/4$ modulation. The signal drops significantly for the $\lambda/2$ modulation case and is nearly independent of the incident laser power as shown in Figure 4b Inset. We believe that this constitutes the background-related portion of our non-local signal. Its origin could be the finite resistance of graphene rather than any laser heating-related artifacts as the signal does not change with increasing laser power. Importantly, the non-local signal under Δλ = 1/4 modulation changes linearly with the laser power. This allows us to achieve a pure spin voltage of ~ 0.35 µV for the laser power of 0.185 mW (Figure 4b-Inset).

Finally yet importantly, we further prove the spin-valley coupling origin of the non-local signal by measuring its dependence on the ellipticity of the modulated light by modifying the incident angle (θ). θ = + (-) 45º indicates the modulation of polarization from R-L (L-R). As shown in Figure 4c, the non-local signal shows very strong dependence on the θ, in a good agreement with the observation in Figure 4b. We observe a maximum signal of 0.6 µV under R-L. The signal decreases and remarkably even changes its sign as the θ is changed. We observe a minimum signal of ~ - 0.2 µV under L-R modulation. The signal for the linear/linear case is ~ 0.3 µV and matches the value obtained for the λ/2 case which was attributed to the background signal. As shown in Supplementary Information, such incident angle dependence is completely absent for λ/2 modulation.

In summary, we have demonstrated the first optical spin injection into graphene by benefiting from the unique spin-valley properties of monolayer WSe$_2$. We activate the spin polarized charge carriers in the WSe$_2$ layer by illuminating the crystal with circularly polarized light. The generated spin current diffuses into the graphene layer and transports over 3.5 µm before its electrical detection through a three layer thick BN tunnel barrier. A recent optical experiment in a graphene/TMDCs based heterostructure suggests that induced charge carriers are electrons[28]. We exclude any spurious effects by prudently studying the separation, power intensity and incident light polarization dependences on non-local signal. In future experiments, Hanle spin precession measurements could be employed to extract the exact spin relaxation length of graphene by applying in-plane magnetic fields. This optically induced spin accumulation could be also detected optically by using Kerr spectroscopy[2,29,30].

**References:**

1. Han, W., Kawakami, R. K., Gmitra, M. & Fabian, J. Graphene spintronics. *Nat. Nanotechnol.* **9,** 794–807 (2014).

2. Gmitra, M. & Fabian, J. Graphene on transition-metal dichalcogenides: A platform for proximity spin-orbit physics and optospintronics. *Phys. Rev. B* **92,** 155403 (2015).

3. Mak, K. F., He, K., Shan, J. & Heinz, T. F. Control of valley polarization in monolayer MoS$_2$ by optical helicity. *Nat. Nanotechnol.* **7,** 494–498 (2012).

4. Žutić, I. & Das Sarma, S. Spintronics: Fundamentals and applications. *Rev. Mod. Phys.* **76,** 323–410 (2004).

5. Wolf, S. A. *et al.* Spintronics: a spin-based electronics vision for the future. *Science* **294,** 1488–95 (2001).

6. Gmitra, M., Konschuh, S., Ertler, C., Ambrosch-Draxl, C. & Fabian, J. Band-structure topologies of graphene: Spin-orbit coupling effects from first principles. *Phys. Rev. B* **80,** 235431 (2009).

7. Trauzettel, B., Bulaev, D. V., Loss, D. & Burkard, G. Spin qubits in graphene quantum dots. *Nat. Phys.* **3,** 192–196 (2007).

8. Mayorov, A. S. *et al.* Micrometer-scale ballistic transport in encapsulated graphene at room temperature. *Nano Lett.* **11,** 2396–9 (2011).

9. Ingla-Aynés, J., Guimarães, M. H. D., Meijerink, R. J., Zomer, P. J. & van Wees, B. J. 24 − μ m spin relaxation length in boron nitride encapsulated bilayer graphene. *Phys. Rev. B*

**92,** 201410 (2015).

10. Schmidt, G., Ferrand, D., Molenkamp, L., Filip, A. & van Wees, B. Fundamental obstacle for electrical spin injection from a ferromagnetic metal into a diffusive semiconductor. *Phys. Rev. B* **62,** R4790–R4793 (2000).

11. Tombros, N., Jozsa, C., Popinciuc, M., Jonkman, H. T. & van Wees, B. J. Electronic spin transport and spin precession in single graphene layers at room temperature. *Nature* **448,** 571–4 (2007).

12. Avsar, A. *et al.* Toward wafer scale fabrication of graphene based spin valve devices. *Nano Lett.* **11,** 2363–8 (2011).

13. Kamalakar, M. V., Dankert, A., Bergsten, J., Ive, T. & Dash, S. P. Enhanced tunnel spin injection into graphene using chemical vapor deposited hexagonal boron nitride. *Sci. Rep.* **4,** 6146 (2014).

14. Li, Y. *et al.* Measurement of the optical dielectric function of monolayer transition-metal dichalcogenides: $MoS_2$, $MoSe_2$, $WS_2$, and $WSe_2$. *Phys. Rev. B* **90,** 205422 (2014).

15. Lopez-Sanchez, O., Lembke, D., Kayci, M., Radenovic, A. & Kis, A. Ultrasensitive photodetectors based on monolayer $MoS_2$. *Nat. Nanotechnol.* **8,** 497–501 (2013).

16. Zeng, H., Dai, J., Yao, W., Xiao, D. & Cui, X. Valley polarization in $MoS_2$ monolayers by optical pumping. *Nat. Nanotechnol.* **7,** 490–493 (2012).

17. Sanchez, O. L., Ovchinnikov, D., Misra, S., Allain, A. & Kis, A. Valley Polarization by Spin Injection in a Light-Emitting van der Waals Heterojunction. *Nano Lett.* **16,** 5792–5797 (2016).

18. Song, X., Xie, S., Kang, K., Park, J. & Sih, V. Long-Lived Hole Spin/Valley Polarization Probed by Kerr Rotation in Monolayer $WSe_2$. *Nano Lett.* **16,** 5010–5014 (2016).

19. Withers, F. *et al.* WSe$_2$ Light-Emitting Tunneling Transistors with Enhanced Brightness at Room Temperature. *Nano Lett.* **15,** 8223–8228 (2015).

20. Zhu, Z. Y., Cheng, Y. C. & Schwingenschlögl, U. Giant spin-orbit-induced spin splitting in two-dimensional transition-metal dichalcogenide semiconductors. *Phys. Rev. B* **84,** 153402 (2011).

21. Muniz, R. A. & Sipe, J. E. All-optical injection of charge, spin, and valley currents in monolayer transition-metal dichalcogenides. *Phys. Rev. B* **91,** 85404 (2015).

22. Jedema, F. J., Heersche, H. B., Filip, A. T., Baselmans, J. J. A. & van Wees, B. J. Electrical detection of spin precession in a metallic mesoscopic spin valve. *Nature* **416,** 713–6 (2002).

23. Takahashi, S. & Maekawa, S. Spin injection and detection in magnetic nanostructures. *Phys. Rev. B* **67,** 52409 (2003).

24. Han, W. *et al.* Tunneling spin injection into single layer graphene. *Phys. Rev. Lett.* **105,** 167202 (2010).

25. Avsar, A. *et al.* Electronic spin transport in dual-gated bilayer graphene. *NPG Asia Mater.* **8,** e274 (2016).

26. Yan, T., Qiao, X., Liu, X., Tan, P. & Zhang, X. Photoluminescence properties and exciton dynamics in monolayer WSe$_2$. *Appl. Phys. Lett.* **105,** 101901 (2014).

27. Tombros, N. *et al.* Anisotropic Spin Relaxation in Graphene. *Phys. Rev. Lett.* **101,** 46601 (2008).

28. Roy, K. *et al.* Graphene-MoS$_2$ hybrid structures for multifunctional photoresponsive memory devices. *Nat. Nanotechnol.* **8,** 826–30 (2013).

29. Yang, L. *et al.* Long-lived nanosecond spin relaxation and spin coherence of electrons in

monolayer $MoS_2$ and $WS_2$. *Nat. Phys.* **11,** 830–834 (2015).

30. Bushong, E. J. *et al.* Imaging Spin Dynamics in Monolayer $WS_2$ by Time-Resolved Kerr Rotation Microscopy. arXiv:1602.03568 (2016).

**Acknowledgements** We thank Kolyo Marinov for his help. A.A., D.U., O.L.S. and A.K. would like to acknowledge support by the European Research Council (ERC, Grant 682332), Swiss National Science Foundation (Grant 153298) and Marie Curie-Sklodowska COFUND (grant 665667). A.K. acknowledges funding from the European Union's Horizon 2020 research and innovation programme under grant agreement No 696656 (Graphene Flagship). B. Ö. would like to acknowledge support by the National Research Foundation, Prime Minister's Office, the Singapore under its Competitive Research Programme (CRP Award No. NRF-CRP9-2011-3), the Singapore National Research Foundation Fellowship awards (RF2008-07), and the SMF-NUS Research Horizons Award 2009-Phase II.

**Authors Contributions.** A.A. and D.U. contributed equally to this work. A.A. and A.K. designed the experiments. A.A. fabricated the samples. A.A., D.U. J.L. and O.L.S. performed the measurements. K.W. and T.T. grew the h-BN crystals. All authors discussed the results. A.A., D.U. and A.K. wrote the manuscript.

**Author Information.** Correspondence and requests for materials should be addressed to A.K. (andras.kis@epfl.ch) and A.A. (ahmet.avsar@epfl.ch)

**Figure Captions.**

**Fig.1 Schematics and device fabrication. a,** Schematics of the device. Electrodes 1, 2 and 3 represent the Co electrodes which have direct contacts to graphene/BN portion of the device.

Electrodes 1 and 2 are used as detector electrodes for junction 1 and 2, respectively and electrode 3 is used as the reference electrode during non-local measurements. Incident beam is focused on WSe$_2$, close to the region at the graphene side. The red spheres with arrows represent the spin generation and diffusion during a non-local spin valve measurement. **b,** Optical image of a typical device at various fabrication stages. WSe$_2$ is transferred onto initially exfoliated graphene stripe. Then, the BN layer is transferred which is followed by the evaporation of the ferromagnetic Co electrodes.

**Fig.2 Electrical characterization of the graphene spin valve with a BN tunnel barrier. a,** Back-gate voltage dependence of graphene resistance. Inset shows the I-V dependence of injector and detector electrodes. They are represented as 1 and 2, respectively in the schematics shown in the inset of Figure 2b. **b,** Non-local signal as a function of in-plane magnetic field. Black and red horizontal arrows represent the magnetic field sweeping directions. Vertical arrows represent the relative magnetization directions of the injector and detector electrodes. Inset: Schematics for non-local spin transport measurement. A charge current of 5 µA is applied from electrode 1 to 2 and the generated spin current is detected by probing the electrochemical potential differences between electrodes 3 and 4. **c,** Hanle precession of the non-local signal as a function of the perpendicularly applied magnetic field. Measurements are performed at $V_{BG} = 0$ V.

**Fig.3 Optical spin injection into graphene a,** Back gate voltage dependence of graphene resistance. Inset shows the I-V dependence of electrodes 1 and 2 which are indicated in the optical image at Figure 3d. **b,** Photoluminescence measurements of monolayer WSe$_2$. **c,** Time dependence of non-local signal while laser spot is moved from graphene to WSe$_2$ and then back

to graphene. Photoelastic modulator is used for enhancing the signal quality. **d,** Non-local signal measured at junction 1 and 2. Junction 1(2) refers to the non-local voltage measured between electrodes 1(2) and 3. Inset shows the optical image of the device. Scale bar is 3 µm.

**Fig.4 Polarization dependence of non-local signal. a,** Spatial map of the non-local voltage signal. The dotted line represents the monolayer graphene, black and blue solid lines represent the monolayer and bilayer $WSe_2$ regions, respectively. The color scale bar is ~ 1.2 µV as it moves from red to blue. **b,** Non-local signal recorded with the incident light under quarter-wave and half-wave modulations. Inset: The dependence of non-local signal on the power of incident light. **c,** Modulation of the spin signal as a function of incident light polarization.

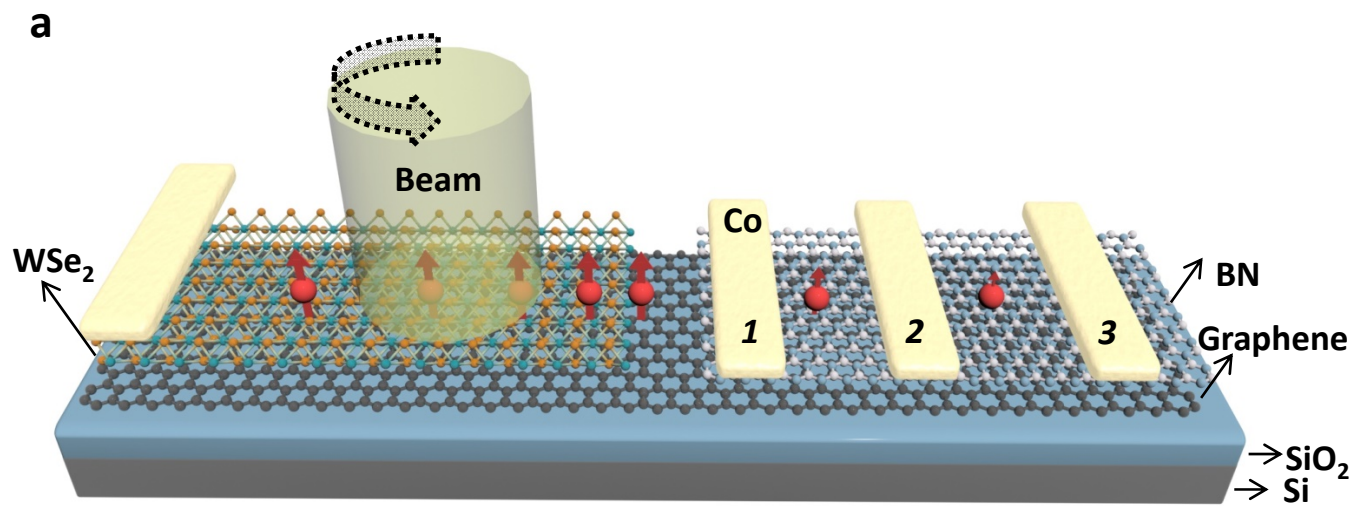
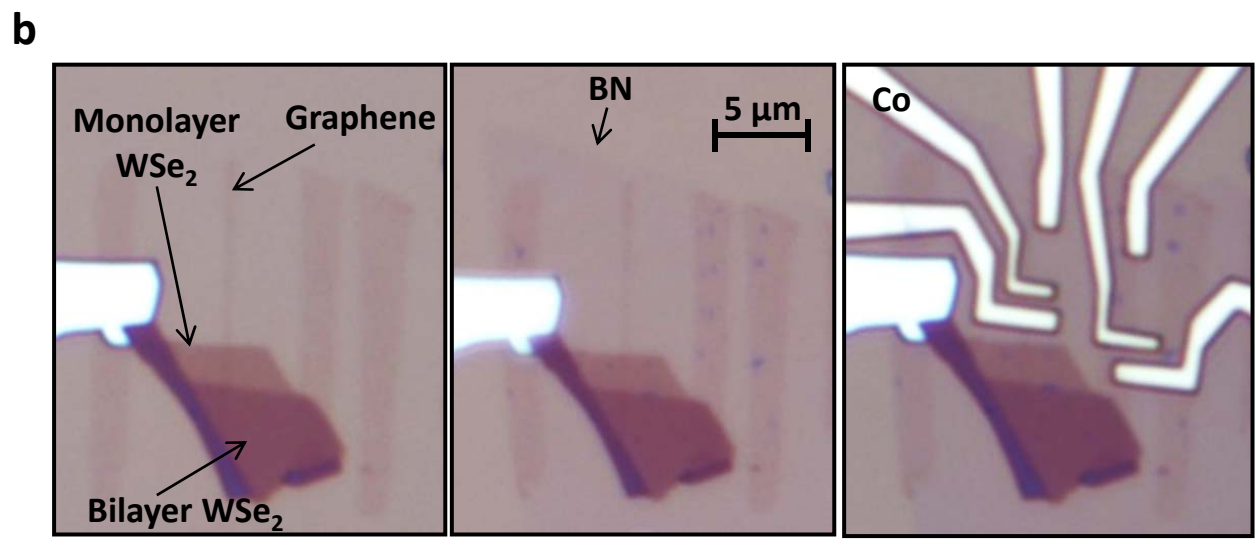

**Figure 1**

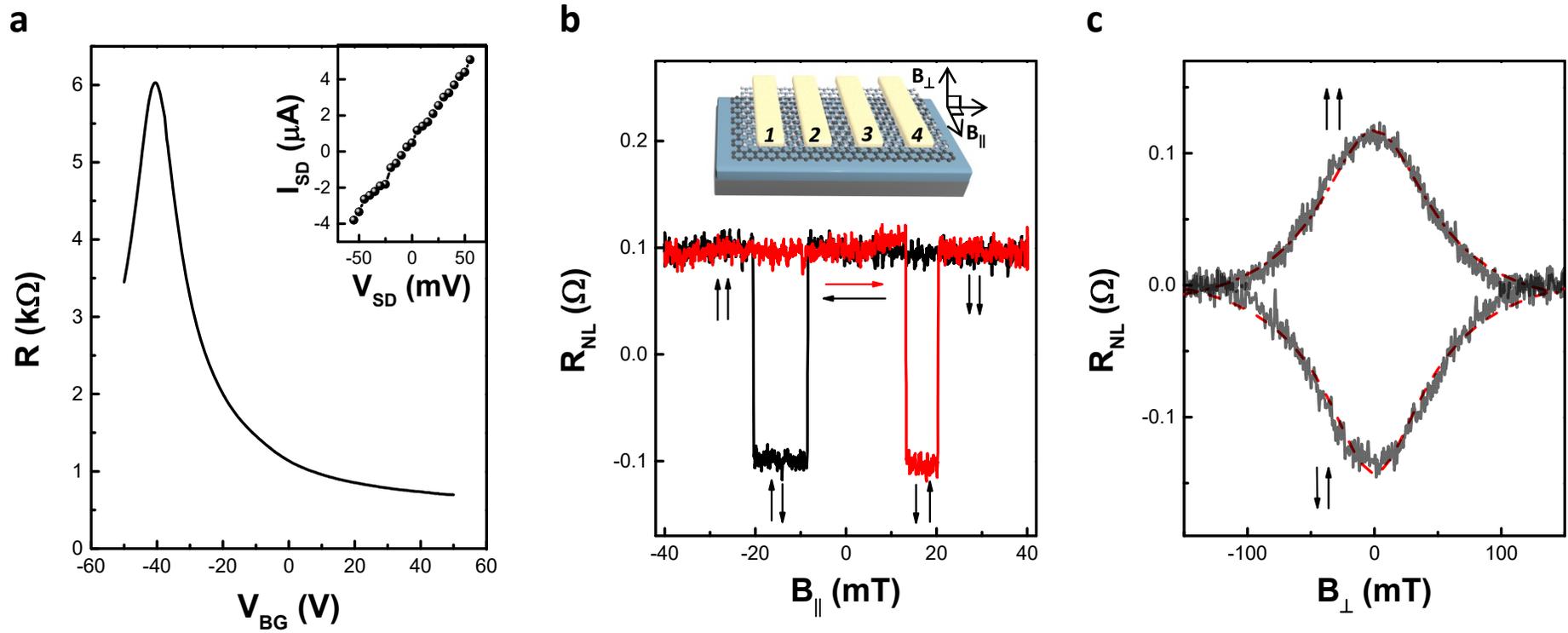

**Figure 2**

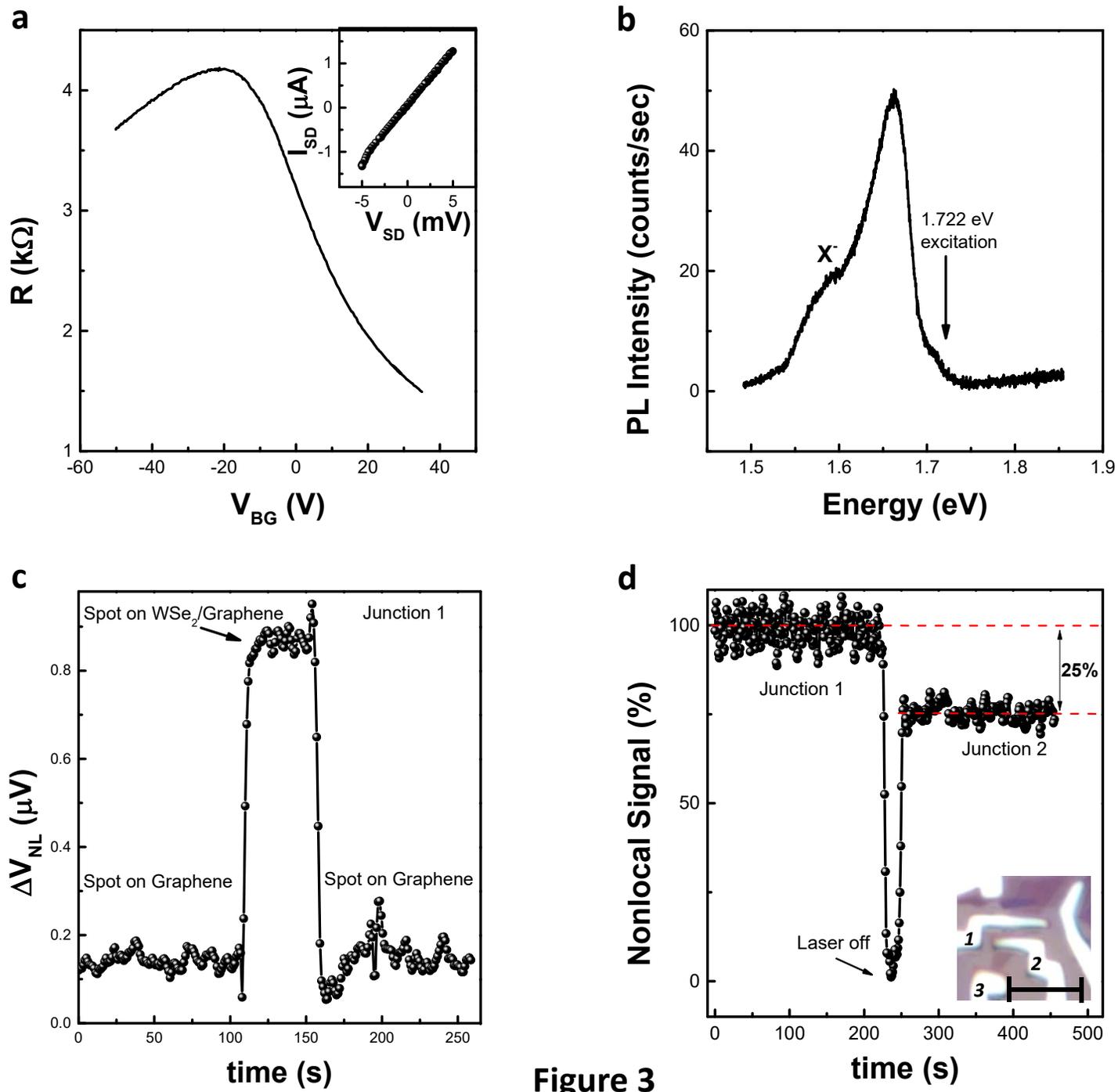

Figure 3

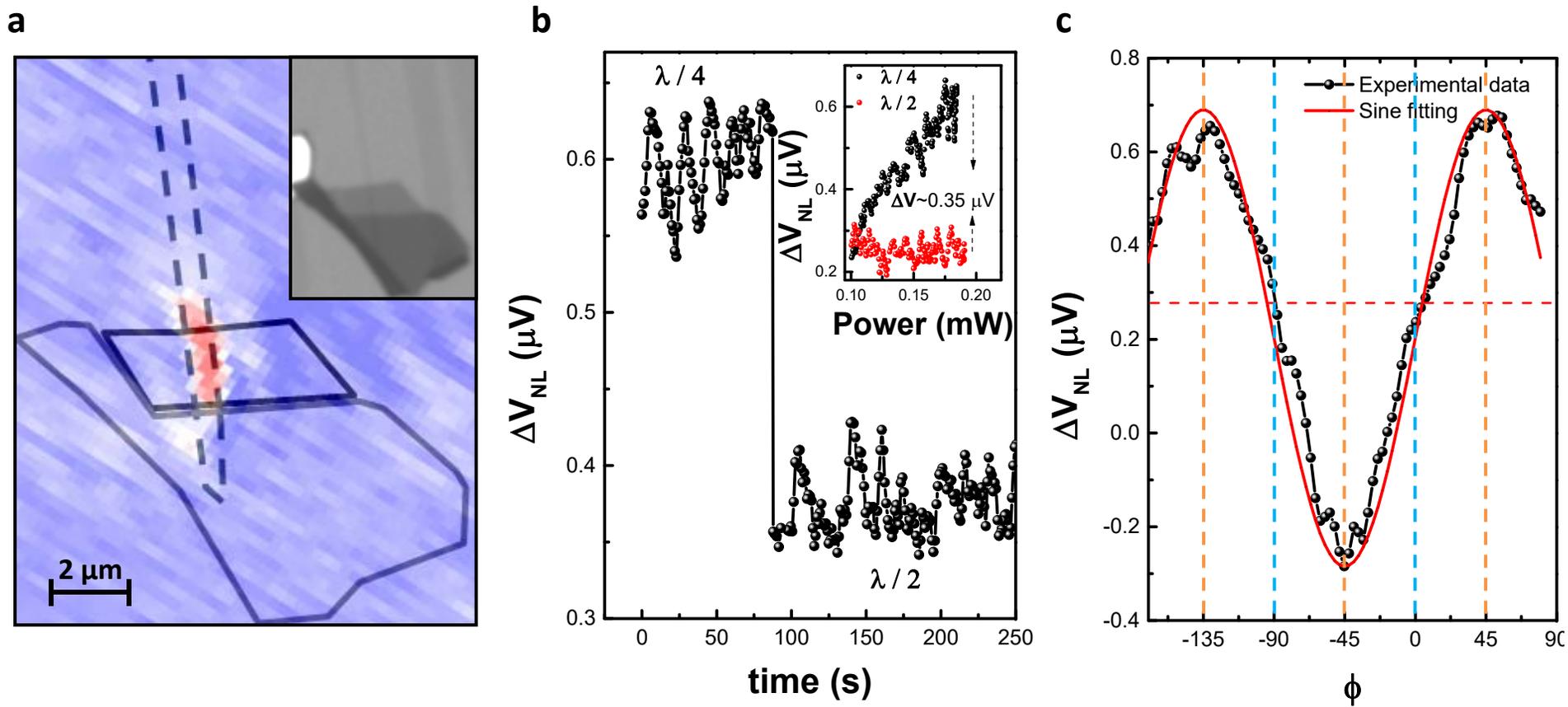

**Figure 4**

# Supplementary Information

# for

**Optospintronics in graphene via proximity coupling**

Ahmet Avsar[1], Dmitrii Unuchek[1], Jiawei Liu[2], Oriol Lopez Sanchez[1], Kenji Watanabe[3], Takashi Taniguchi[3], Barbaros Özyilmaz[2], Andras Kis[1]

[1] Electrical Engineering Institute and Institute of Materials Science and Engineering, École Polytechnique Fédérale de Lausanne (EPFL), Lausanne CH 1015, Switzerland

[2] Centre for Advanced 2D Materials, National University of Singapore, Singapore 117542, Singapore

[3] National Institute for Materials Science, 1-1 Namiki, Tsukuba 305-0044, Japan

## The PDF file contains:

1-3 Schematics for optospintronics measurement configuration.

4. Gate voltage dependence of extracted spin transport properties.

5. Photoluminescence spectra of $WSe_2$.

6. Spot location dependence of non-local signal.

7. Incident angle dependences of non-local signal for $\lambda/2$ and $\lambda/4$ modulations.

8. Gate voltage dependence of non-local signal.

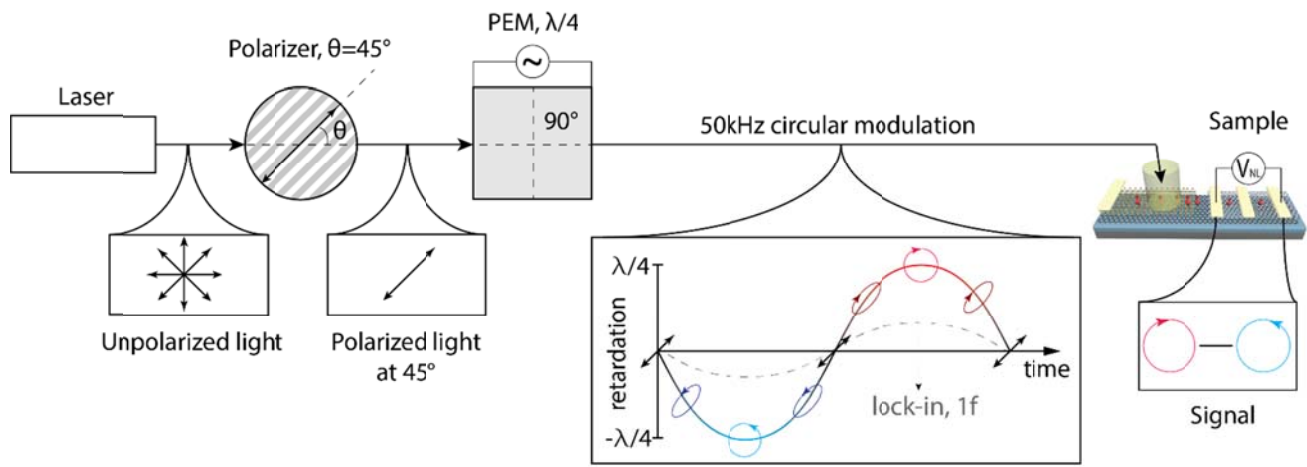

**Figure S1│ Quarter wave (λ/4) modulation by PEM.** We polarize the initially unpolarized light by the linear polarizer oriented 45° to the optical axis of the photoelastic modulator (PEM). This orientation of the incident light provides the highest degree of modulated light circularity. PEM acts as a variable birefringent plate providing time-dependent retardation along one of the axes at frequency of 50 kHz (1f). In the case of λ/4 modulation, applied retardation has maximal (minimal) value of λ/4 (-λ/4) with the PEM acting as a quarter wave plate at these moments, thus generating the right (left) circularly polarized light. Lock-in amplification of the non-local signal with the PEM (1f) fundamental frequency (50 kHz, gray dashed line) results in a signal that corresponds to the variation of the non-local signal caused by the right and left-handed light. Therefore, the resulting light modulation is right-to-left (left-to-right) in the case of 45° (-45°) incidence angle.

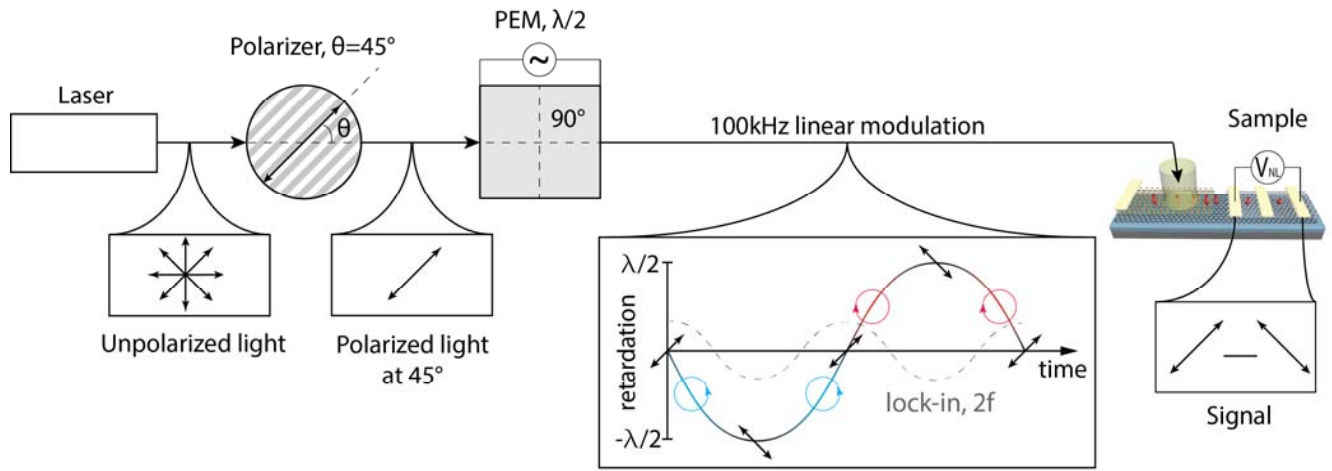

**Figure S2│ Half wave (λ/2) modulation by PEM.** In the case of λ/2 modulation, applied retardation has maximal (minimal) value of λ/2 (-λ/2) with PEM acting as a half wave plate at these moments, thus rotating the incident polarization by 2θ (90° in the case of 45° incidence), preserving its linearity. Lock-in amplification of the sample electrical response with the reference signal (gray dashed line) at double the PEM operating frequency (100kHz, 2f) results in a signal that corresponds to the variation in the non-local signal caused by the original and rotated linearly polarized light. Here, the resulting light modulation is linear to linear.

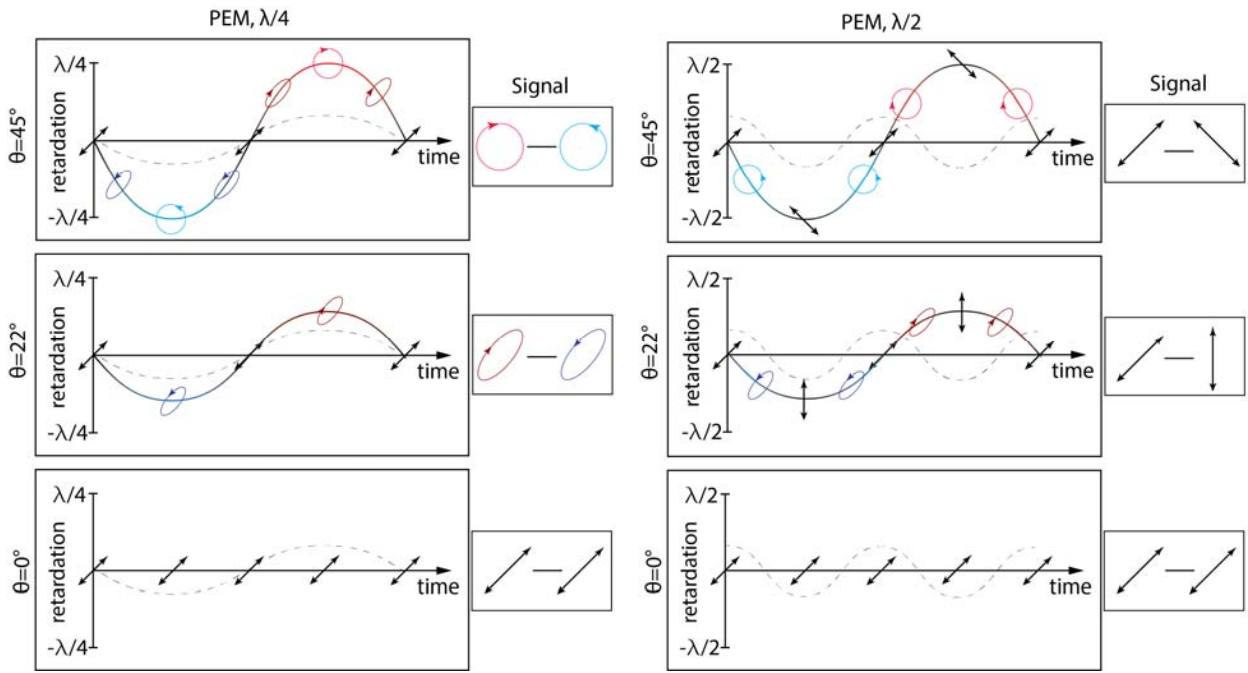

**Figure S3│ Light modulation dependence on the polarization angle rotation.** This sketch represents schematic results of light modulation for three particular angles of incidence (θ = 45°, 22°, 0° from the top to the bottom) in the cases of quarter wave and half wave modulation (left and right parts respectively). Since the PEM operated in a λ/4 mode acts as a time-dependent quarter wave plate, it results in the modulation of the circular handedness of the light if incident at 45° angle to the PEM axis (upper left drawing). Deviating θ from the 45° angle, results in the components of light parallel and perpendicular to the PEM optical axis being taken out of balance, thereby increasing ellipticity of the modulated light (middle left). Light incident at θ = 0° (bottom left schematic) is not influenced by the PEM since it is oriented along its optical axis while only the component across this axis gets retarded. In other words, angle θ has a direct impact on the ellipticity degree of the light modulated by PEM operating in λ/4 mode and thus on the signal amplified by the lock-in. On the other hand, half-wave modulation (right part) is fundamentally different from the former case, since light is modulated linearly between θ and – θ at twice the frequency of PEM operation.

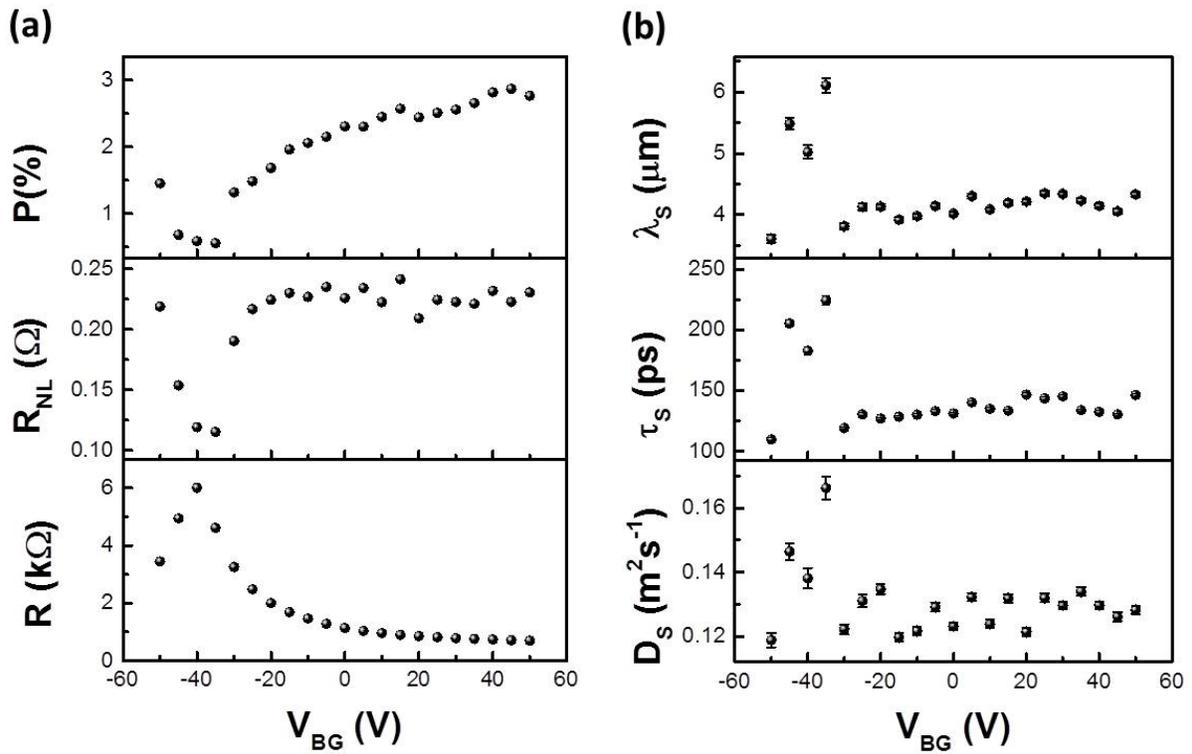

**Figure S4│ Gate voltage dependence of the extracted spin transport properties. (A)** Back gate voltage ($V_{BG}$) dependences of spin polarization (P), non-local resistance ($R_{NL}$) and local resistance (R). **(B)** $V_{BG}$ dependences of spin relaxation length ($\lambda_S$), spin relaxation time ($\tau_S$) and spin diffusion constant ($D_S$).

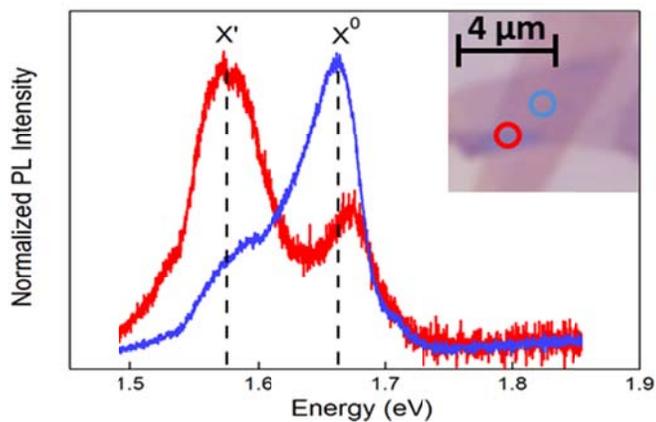

**Figure S5 | Normalized photoluminescence spectra of WSe$_2$.** Blue curve represents the PL spectrum obtained at the center of the WSe$_2$ monolayer, while red curve shows PL from the more defective edge of the flake. In the center of the monolayer, the excitonic emission is dominant (1.66eV, X$^0$). In contrast, higher density of defects as well as the presence of bilayer in the edge area results in stronger recombination of the localized electrons leading appearance of well-pronounced low energy peak (1.58eV, X'). Insert shows the optical image of the device with indicated locations of PL measurements. All PL spectra were obtained at 488nm excitation wavelength with 40uW incident power. Measurements were performed at 4K.

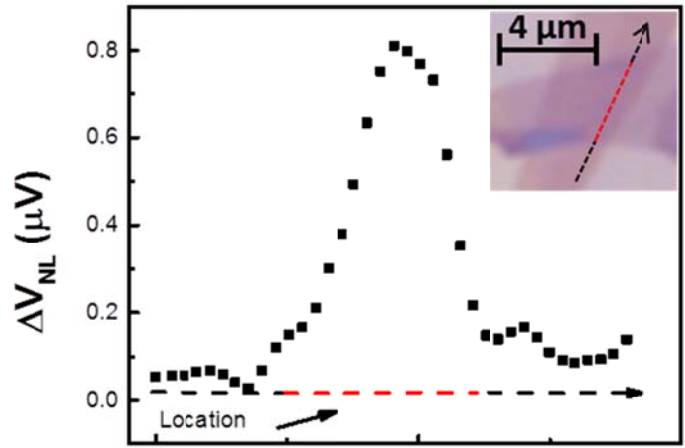

**Figure S6│Location dependence on non-local signal.** We move the incident light spot across the heterostructure as the non-local voltage is recorded for the case of λ/4 modulation. Excitation wavelength is 720 nm and incident power is 180uW. Red arrow shows the direction of motion. We see an enhanced signal once the spot is on the WSe$_2$/graphene interface but the signal is slightly location dependent. Measurements were performed at 4K.

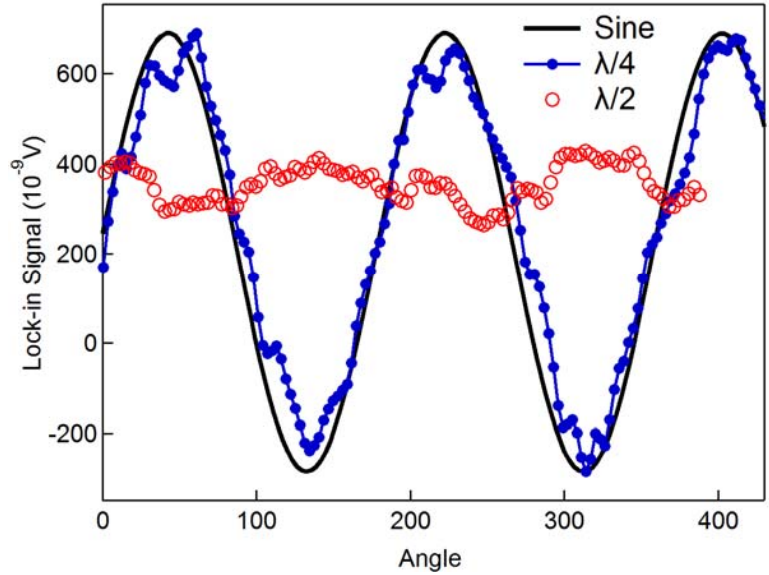

**Figure S7│ Spin signal dependence on the ellipticity degree.** Non-local signal as a function of the incident light angle θ at half-wave modulation (red dots) and quarter-wave modulation (blue dots). Black solid line shows sine function of twice the angle of incidence (2θ) as a representation of the ellipticity degree of the modulated light.

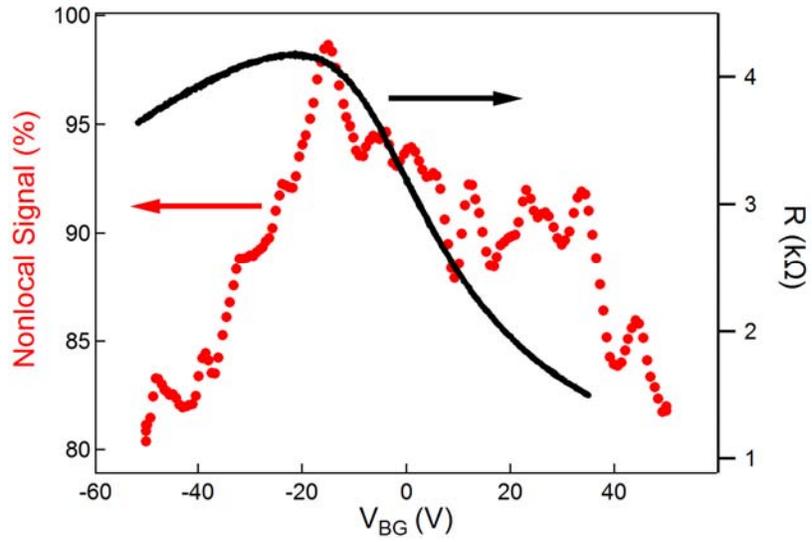

**Figure S8│Nonlocal signal as a function of back-gate voltage $V_{BG}$.** Spin signal dependence on the applied gate voltage (red points) follows the resistivity of the graphene channel (black line). Maximum signal at $V_{BG}$ = - 20V matches the Dirac point of the graphene. Electrostatic doping decreases both the resistivity of the channel and the non-local signal generated through the optical spin injection scheme.